# IMPROVED QUALITY OF SERVICE PROTOCOL FOR REAL TIME TRAFFIC IN MANET

Iftikhar Ahmad[1], Humaira Jabeen[2] and Faisal Riaz[3]

[1]Department of CS & IT Mirpur University of Science and Technology
Mirpur, AJK, Pakistan
`ify_ia@yahoo.com`
[2]Department of Computer Science International Islamic University
Islamabad, Pakistan
`humaira_jabeen_83@yahoo.com`
[3]Department of Computing and Technology, Iqra University Islamabad Campus
H9/1 Islamabad, Pakistan
`fazi_ajku@yahoo.com`

## ABSTRACT

*The technologies like Wi-Fi, Blue tooth, WiMax etc. have made Mobile Ad hoc Networks common in our Real life. Multi-media applications need to be supported on MANET. A certain level of QoS (Quality of Service) support is essential for Real time data. Our proposed protocol provides the required QoS without having negative impact on Best Effort data traffic. An efficient rout discovery mechanism for AODV routing protocol as well as transmission technique for real time data are proposed. This technique gives more transmission opportunities to real time data traffic results in decreasing transmission delay and increasing throughput. A modified version of the popular AODV routing protocol to provide QoS guarantee for real time traffic in MANETs is proposed. The simulation shows better performance results for proposed protocol over the basic AODV.*
.

## 1. INTRODUCTION

A mobile ad hoc network (MANET) [1] is a wireless network of autonomous mobile nodes, the network having no infrastructure and runs on batteries. Communication occurs between nodes without a central entity like a base station or an access point. This allows the mobile nodes to setup cost effective networks quickly and without any fixed topology. MANET is originally designed for use in military applications, but can also play an important role in relief and rescue operations. Most recently, the application of MANETs in innovative paradigms such as Vehicle-To-Vehicle communication and wireless sensor networks has also been explored. These networks are primarily made to transmit quick information exchange between the participating nodes; the existing routing protocols supports best effort type of traffic. The MANET protocols cannot guarantee reliability and QoS because they contain self-configured nodes with a dynamic network topology. The dynamic nature of the network topology is mainly attributed to the mobility of the nodes. Maintaining accurate information about the status of a network for routing purposes becomes difficult because of the constraints of the Mobile Ad hoc Network.

MANETs employ two basic types of routing protocols, namely table-driven protocols and on-demand routing protocols. The table-driven protocols are proactive protocols that consume network bandwidth but have less routing overhead. On the other hand, the on demand routing protocols are reactive protocols that exchange routing information only when needed. The on demand protocols are often preferred in bandwidth constrained environments. Among the available on demand protocols, the Ad-hoc On-demand Distance Vector (AODV) is most widely used because of its simplicity, scalability, low computational complexity and low overhead. AODV routes IP packets to the destination node by discovering the shortest path without giving due consideration to network reliability and QoS [2]. This fundamental drawback limits the scope of AODV as it cannot perform well for real time applications, which require certain QoS guarantees like fixed bandwidth, low latency and jitter, less link failures, etc. As mentioned earlier, these guarantees cannot be accommodated in the conventional AODV due to the inherently dynamic nature of MANETs.

The conventional basic AODV discovers the route by broadcasting route request (RREQ) messages throughout the network. Based on the responses received from the participating nodes, the route with least number of hops to destination is selected for data transmission. Modified versions of AODV also consider factors like traffic load on a route, link quality among nodes, energy consumption, traffic flow type etc. This route discovery process is initiated when a new node joins the network or when a link between the neighboring nodes breaks down. A node first stores an incoming packet in its buffer and executes a queuing mechanism to place the stored data packets on the transmission channel. The delay in a network increases with the size of the queue in a buffer. If the length of a queue exceeds a certain threshold, packets are dropped. Traffic congestion occurs when the rate of packet arrival in the buffer exceeds the rate of transmission. Queuing delay, packet loss and traffic congestion should be controlled in order to support the real time applications. This paper is extended version of QoS AODV routing protocol in to provide QoS-based routing in MANETs.

A mechanism is proposed to provide QoS guarantee to support real time applications. More scheduling priority is given to the data packets originating from real time applications.

The rest of this paper is organized as follows. Related work and problem definition is presented in Section II. Best Effort Traffic aware rout discovery and packet forwarding mechanism is covered in Section III, followed by implementation results and performance evaluation in Section IV. This paper is concluded in Section V while the references are given towards the end of paper.

## 2. RELATED WORK

The Quality of Service based routing services has been presented to a greater extent in the previous works. The work is done to meet the QoS needs of multimedia applications over the MANET.

FQMM [3] has been presented as the first QoS model for MANET which is a hybrid of both Integrated Service and Differentiated Services architectures. Salient features of FQMM include: dynamics roles of nodes, hybrid provisioning and adaptive conditioning. The Ad hoc QoS on-demand routing (AQOR) protocol has been proposed in [4] that deals with the bandwidth and end to end delay requirements. On demand route discovery, signaling function and hop to hop routing are the main components of the proposed protocol. But the protocol does not deal with the queuing delay that is main requirement of the QoS model for real time applications. More recent works on QoS issue are summarized in the following.
An on demand delay based quality of service (QoS) routing protocol (AODV-D) protocol [5] is proposed for QoS routing to ensure that delay does not exceed a maximum value for the MANETs. MAC layer channel contention information and number of packets in the interface queue are considered in addition to minimum hop criteria for route discovery in MANETs. AODV-D focuses on reducing the network delay but does not consider the bandwidth requirements of real time applications. A QoS routing algorithm, called QoS AODV (QAODV) has been proposed in [6] as an extension to AODV-D. A weighted function of several parameters is used to select optimal routes and to provide support for QoS and fault tolerance. QAODV has also shown advantages in terms of throughput and delay. It has limited support for multimedia applications.

QEAODV routing protocol [7] improves the normal route finding method of AODV for providing QoS in MANETs. QEAODV establishes a path between the source and the destination on the basis of meeting the application throughput requirement. QEAODV handles the channel access contention effectively which is the inherent problem in MANET. Cluster based QoS routing (CBQR) [8] is a table driven routing protocol that provides support for bandwidth efficiency and not only deals with the bandwidth requirement over the wireless network but also takes care regarding the stale routes, storage overheads and limited battery power. Another QoS routing protocol (MDSR-AODV) for MANETs with Link Stability Model is proposed in [9]. MDSR-AODV makes routing decisions by taking into account the characteristics of the channel. It finds paths with greater link stability factor in the route discovery phase. A long-lived path is preferred for data transfer among the available options. by monitoring network topology changes through delay prediction the route maintenance is done.

Dynamic Load-Aware Routing (DLAR) [10] protocol defines the network load of a mobile node as the number of packets in its interface queue. Load-Balanced Ad hoc Routing (LBAR) [11] protocol define network load in a node as the total number of routes passing through the node and its neighbors. In [12] Load-Sensitive Routing (LSR) protocol the network load in a node is defined as the summation of the number of packets being queued in the interface of the mobile host and its neighboring hosts. Even though the load metric of LSR is more accurate than those of DLAR or LBAR, but it does not consider the effect of access contentions in the MAC layer .Therefore, LSR produce contention delay that increase overall delay of transmission.

The contention delay problem is minimized in Delay-Oriented Shortest Path Routing protocol [13]. The protocol analyzed the medium access delay of a mobile node in IEEE 802.11 wireless network.

In [14] a protocol is proposed which includes load parameter in the RREQ message that helps in the selection of route with low congestion during route discovery process. This decreases the end to end delay for real time traffic. Route maintenance is done through RODV by not sending the error information back to the source results a decrease in routing overhead.

Aforementioned protocols used different techniques to increase the performance of the network. We have seen different load and delay reduction methods in the literature. Number of techniques for real time data transmission is proposed as well. Most interesting idea in QoS provision is by adding messages during the route discovery phase that carry QoS requirements. These message extensions are mostly about bandwidth requirements and delay threshold value. Almost every protocol in HWN and MANET is fully or partially adopting the technique of adding extension to RREQ message in order to provide QoS guarantees to the delay sensitive applications But we have proposed new route discovery mechanism that not only taking care of overall traffic load on the rout but specially Best Effort traffic status of the selected rout. New rout discovery mechanism is suitable for real time multimedia traffic. In order to reduce overall delay a scheme is proposed to minimize queuing delay without effecting Best Effort traffic on the network. The literature review gives us an insight of different routing approaches that has been made for MANET to achieve optimal performance for real time data transmissions in the given network scenarios. After deep literature understanding it is quite clear that because of the different nature of MANET, not any single routing protocol work well with all network scenarios of the MANET.

# 3. RELATED WORK

In our proposed solution we use reactive routing protocol Ad hoc on Demand Distant Vector Routing (AODV) for routing in Mobile ad hoc network. We have proposed two schemes, firstly we modified existing rout discovery mechanism and secondly the packet forwarding technique.

## 3.1. Balanced Best Effort Traffic Aware Route Discovery

When a source node wants to communicate with another node for which it has no routing information available. The route discovery process starts and Source node broadcasts RREQ message to its neighbors. The RREQ message contains a reserved bits field in it. This field is utilized to carries the load status information of each node in the rout. When a node receives the RREQ message, the node will add value to the value present in RREQ message's reserved field. At each node it is counted that how many number of Best Effort (BE) packets are there among other type of packets in the queue (Buffer). At each node the number of BE packets in the queue of that node is counted and then ratio of number of BE packets to the number of rest of the packets of queue is calculated. Finally this ratio is added to the reserved field of the RREQ message as shown in figure 3 above.

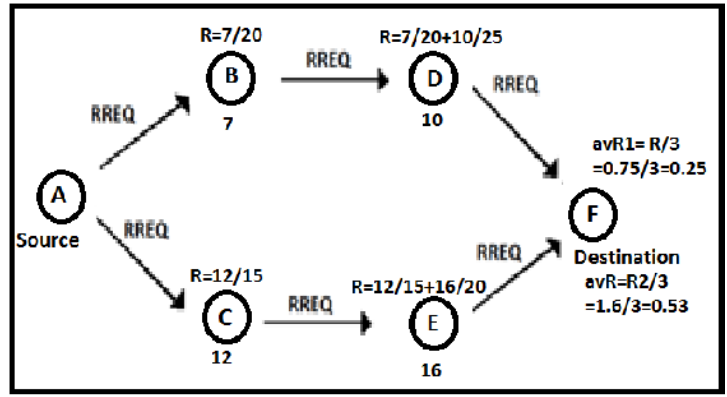

Figure 1. Route request process

In the figure above the value below the node is indicating the number of BE packets in the queue of that node. The R is the ratio of BE packets to the rest of the packets. The value of R above the node is showing the addition of ratio to the reserved field.

This process is repeated at each node through which the RREQ message is passed and reached up to the destination. The value of R is finally reached at the destination and the destination will divide R with hop count value of the REEQ message it has received and produces the average R. The destination will compare the average R value with all requests it received with same sequence number and broad cast ID and unicast RREP message back to the neighbor whose average R value is less. Figure 3 above is explaining the whole process in detail.

In this way the destination will receive two RREQ messages with the same number of hop count is 2 but with two different average R value. Now the destination will select the optimal path on basis of average R on the nodes. Destination will compare both R, s value, higher R value means more number of packets in the rout, and lesser value means less number of BE packets in the route. Hence, the path shown below will be selected as the route from source to the destination. The Figure 4 is also displaying the rout reply path that is finally selected for transmission.

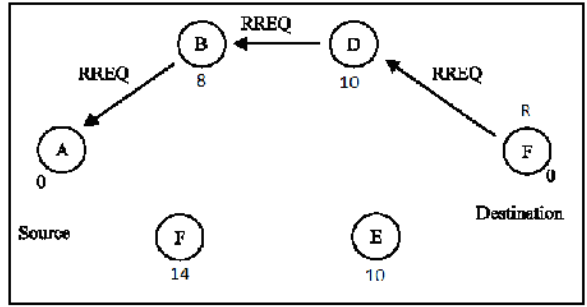

Figure 2. Route reply through selected path

Actually, the destination will selects one of the RREQ messages and ignore others for route reply on the basis of reserved value and it selects node D as in above figure 4. The destination nodes F will unicast the RREP message to the source in the selected optimal path. By selecting routes in this way we are able to overcome the limitation that the nodes with more BE packets load will come in the active route. This will prevent the node with more BE traffic loads, not to take part in the route if another route is available. We are selecting the path with nodes having fewer BE traffic loads, the BE traffic will not be effected badly when our new packet forwarding procedure is adopted and real time data is given priority.

## 3.2. Packet Forwarding Procedure

For forwarding of real time packets we modified existing packet scheduling mechanism of the protocol. Normally packets those are waiting to be forwarded towards destination, both best effort and real time are, buffered or stored in a queue on source node as well as on each intermediate node. Without considering that whether it is a real time or a best effort packet, packets are sent from queue one by one as they come to the head of the queue.

Source node first checks its sending buffer for packet of particular destination whenever source node wants to communicate with destination node. The source can also act as intermediate node for flows coming from other nodes so they can be best effort packets, real time packets and control packets as well. Our proposed mechanism for QoS for real time traffic has main idea that the real time packets resides for minimum time in Queue. Normally, source node checks packets in sending buffer for a particular destination in the network. If node found packet in the queue then simply forwards it one by one without checking its type. In our proposed solution we are treating best effort packets and real time packets separately. Each node in the route first checks its sending buffer for a packet of particular destination and at the same time the type of packet as well. If the packet is RT then it is transmitted immediately.

For a larger time only real time packets from queue are transmitted in order to give more priority to real time packets. A lesser time slot is reserved for BE packets. In this way more transmission opportunity is given to RT data packets than BE data packets. The same procedure will be followed by all the intermediate nodes in the route from source to the destination. By following this procedure the destination will receive more real time packets in less time than normal scheduling procedure. Throughput for real time packets is increased. The following paragraphs describe the packet forwarding procedure in detail.

| Packet Position | 1 | 2 | 3 | 4 | 5 | 6 | 7 | 8 | 9 | 10 |
|---|---|---|---|---|---|---|---|---|---|---|
| Packet Type | RT | BE | BE | RT | RT | BE | BE | BE | RT | RT |

Table 1. Packet Positions in the queue

Table 1 is displaying an example queue and respective positions of Real time (RT) and Best Effort packets (BE). Packet position shows position of packet inside the queue. Packet position 1 represents Head of the queue and packet position 10 represents Tail of the queue. Packet type shows type of packet i.e. either it is BE packet or RT packet. We assumed that maximum of ten packets can be stored at any given time in the queue.

If one packet can be forwarded to the destination in 1ms then 5 packets can be transmitted in 5ms. According to the normal scheduling procedure two real time and three best effort packets forwarded to destination in first five seconds.

In such case the destination for real time traffic receives 2 packets in 5ms. When the load of best effort traffic on the network increases then the number of real time packets forwarded to the destination decreases by adopting normal procedure. Because the real time packets will get stock in the queue and have to wait longer in the queue. As a result the throughput of real time traffic becomes low that is not required QoS for real time traffic.

We proposed a new procedure when the node checks for the packets for a particular destination at the same time node checks the type of packet and in first 5ms five real packets forwarded to destination. More time slot for real time traffic and lesser wait in the queue which are the required parameters for the real time traffic. So the destination receives more real time packets in less time which increases the throughput of real time traffic.

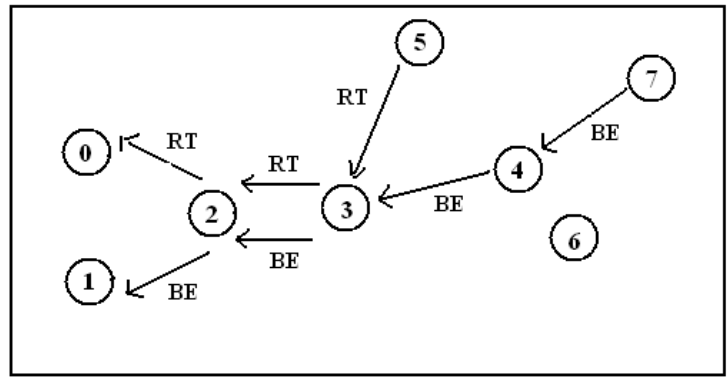

Figure 3. Transmission of BE and RT traffic through network

For specific time period node search for real time packets only that means more transmission opportunity is given to real time packets. The time slot for the real time traffic is reserved in a sense that for a specific period of time the node checks only for real time packets in the queue. If any RT packet is there during that time slot they will be sent immediately towards the destination and node will never care for other type of packets at all during that particular time. Consequently, a smaller time slot is reserved for BE packets. The proposed method is quite opposite to the method of weighted fair queuing. In weighted fair queuing method each data flow is given fair chance of transmission regardless of their packet type. Instead of fair chance we are giving more transmission opportunity to the RT traffic then BE traffic.

The proposed procedure not only reserves more bandwidth to the real time traffic, it also reduces queuing delay for RT packets. RT packets do not have to wait in the queue for longer time because now RT packets are treated on priority basis, which were normally treated on first come first serve basis. In first come first serve basis if a packet is on 9th position in the queue it will have to wait for the eight packets to be transmitted. Now if the packet is RT then the position in the queue will never matter for its transmission.

As shown in figure 1 mobile node 7 is moving and sending packets towards node 1 through node 3 and node 4. Node 5 is transmitting to node 0 through the path 5→3→2→0 as well. Node 5 generates only RT traffic and node 7 generates only (BE) best effort traffic. Both node 5 and node 7 can send both types of data packets simultaneously but for simplicity we have implemented

| Packet Position | 1  | 2  | 3  | 4  | 5  | 6  | 7  | 8  | 9  | 10 |
|-----------------|----|----|----|----|----|----|----|----|----|----|
| Packet Type     | RT | BE | BE | RT | RT | BE | BE | BE | RT | RT |

likewise as in figure 1.

Table 2. Status of queue after new procedure

This thing makes statistics and calculation easier. Both type of traffic RT and BE passing through node 3 and node 2. The queue status is shown in the figure 2 at a specific time interval during the transmission.

Now with the new packet forwarding algorithm if in 1ms queue forwards one packet then during 8ms the queue sent 8 packets to the destinations. Three BE packets and five RT packets but during first 5ms five RT packets from position 10, 9, 5, 4 and 1 are taken for forwarding.

RT packets are transmitted immediately, normally we can see that the packet at position 1 would have to wait for 9ms and packets at 4th and 5th would have to for 5ms and 6ms. These times delays are reduced by forwarding them according to new procedure. BE traffic is given low bandwidth as they are required less bandwidth.

We gave more transmission opportunity to real time data packets as RT data packets require more bandwidth and assured forwarding. Possibility of dropping of real time data packets in case of congestion on the node in network is less. We gave less transmission opportunity to best effort packets for routing as best effort packets requires no tight bound for its forwarding, through the network. Hence, delay is minimized which results in to more throughput for real time traffic. One thing is noticeable that BE traffic is suffered in certain extent. In order to overcome this problem we did two things; one we have proposed new rout discovery mechanism in which a rout is selected that have lesser BE effort traffic on it. Secondly, a certain transmission opportunity is also given to the BE traffic.

International Journal of Computer Networks & Communications (IJCNC) Vol.5, No.4, July 2013

## 4. IMPLEMENTATION AND ANALYSIS

Papers We implemented our improved AODV QoS based in NS2 [15] and compare its performance with basic AODV. Propagation model is Two Ray Ground and Mobility model is Random Way Point [16] because these models are most widely implemented in simulations of most literature work.

### 4.1. Traffic Generation

Two traffic generation applications in our simulation process are CBR and TCP. CBR to simulate Real time flows, whereas TCP connection that simulates FTP to simulate best effort traffic. TCP generate BE traffic that does not require any service quality during transmission.

### 4.2. NS-2Parameter Setting

Other parameters of NS2 are configured for our network scenario is shown in the Table 1. NS2 is discrete network simulator that is used for implementation for performance comparison MANET protocols.

Table 3. NS-2 Parameter Setting

| X dimension of topology | 700 |
|---|---|
| Y dimension of Topology | 600 |
| Routing protocol | AODV |
| Link layer type | LL |
| MAC type | Mac/802_11 |
| Interface Queue | Queue/Drop Tail/PriQueue |
| Maximum packets in Ifq | 20 |
| Antenna model | Antenna/Omni Antenna |
| Radio propagation model | Propagation/TwoRay Ground |
| Network Interface | Phy/WirelessPhy |
| Channel type | Channel/Wireless Channel |
| Number of mobile nodes | 15 |
| Simulation time | 40 Seconds |

### 4.3. Performance Metric

We compared our improved AODV protocol for real time applications with basic AODV without scheduling mechanism. We evaluated the performance according to following metric for RT traffic.

82

**4.3.1. Average Throughput**

Throughput is the total number of RT packets received successfully in the given time at the destination node.

**4.3.2. Transmission Delay**

Transmission delay includes queuing delay and overall transmission delay from source to destination for real time traffic. In our case queuing delay is reduced which reduced overall transmission delay.

**4.4. Simulation Results**

The Basic AODV is first implemented and run on given MANET scenario in NS-2 and performance parameters are calculated. Results are generated in graphical form produced by using trace files that is generated during simulation. Then our improved protocol for real time traffic is implemented and results are produced in graphical form.

Results are compared by taking parameters (pause time) on X-axis and performance metric throughput and delay on Y-axis. In the figure below at equal intervals the throughput is calculated as shown in figures.

**4.5. Comparison of Performance**

Normally, throughput of RT packets is much less then throughput of BE packets as in figure 2. At start throughput of both real time and best effort packets are equal but after that throughput of BE packets increased significantly.

The difference of performance between basic AODV and our improved QoS AODV for real time traffic with new rout discovery and packet forwarding procedure is clear as shown in figure 4.Throughput of our proposed protocol for real time packets is higher than the throughput of basic AODV for real time. The overall delay of AODV is much larger then overall delay of our proposed protocol, which is critical for real time application over MANET.

Now the comparison is performed only for real time packets to check and validate the performance of our proposed protocol for real time transmission. The comparison is performed between throughputs of real time packets reaches at destination.

In figure 4 and figure 5 shows throughput of RT packets, before and after the improved AODV. In the start of transmission at t=5 sec there is no major difference between the two throughputs but as the pause time increases, difference between two throughputs significantly increased.

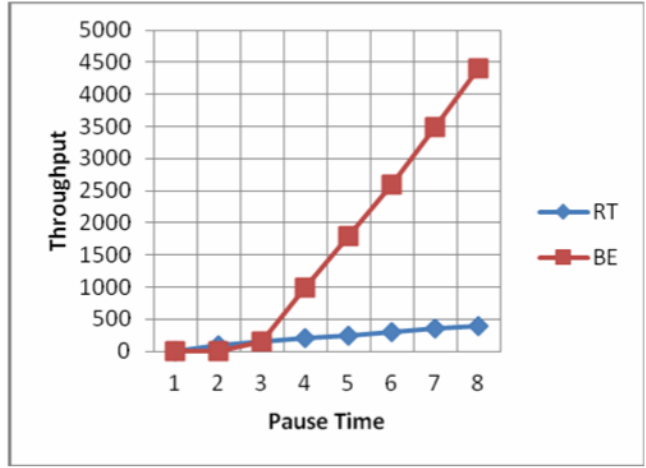

Figure 4. Throughput Vs Pause time without improvement

In case of basic AODV the total of 4700 of BE packets and 450 RT packets reached at the destination as shown in figure 4. But in case of our improved AODV the total of 3500 BE packets and 1550 RT packets successfully reached the destination as shown in figure 5. A greater improvement is clearly displayed for RT transmission with a little decrease in performance for BE transmission.

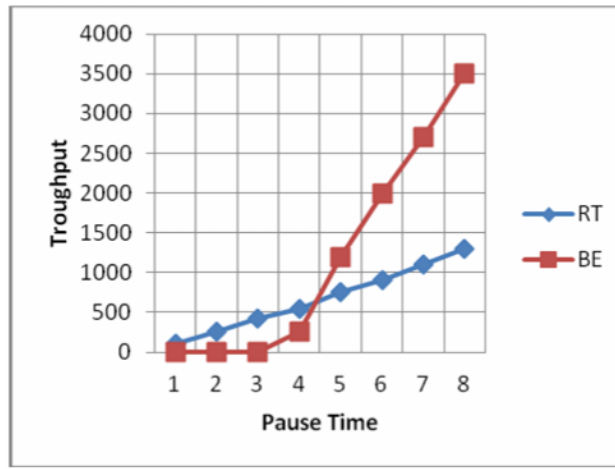

Figure 5. Throughput Vs Pause time of improved AODV

The figure 6 displays the throughput of basic AODV for RT transmission named as Normal Throughput versus throughput of improved AODV named as RT Throughput. Our improvedQoS AODV protocol shows greater improvements as indicated by blue line in figure 6.

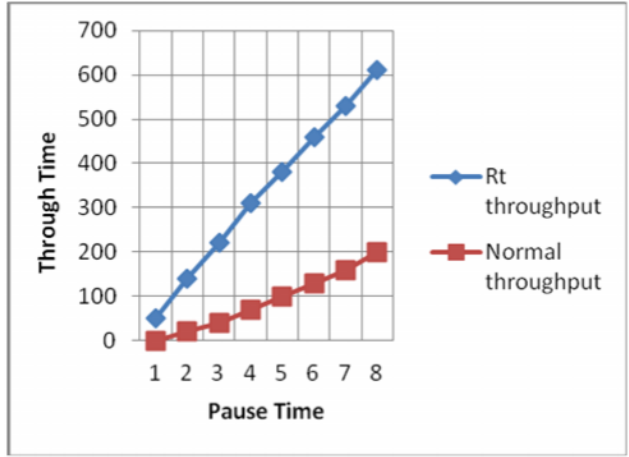

Figure 6. Throughput Vs pause time before and after improved AODV protocol

The figure 7 displays the delays of real time traffic, the blue line shows that the delay of improved AODV for RT traffic and red line shows the delay of basic AODV. Our improved AODV protocol decreases queuing delay and overall delay as well which further minimizes overall delay for RT traffic. The transmission delay of improved AODV is much lower than the normal delay of RT packets with basic AODV.

In case of basic AODV each RT packet takes round about 0.2ms on average. It means in 0.2ms one packet reaches successfully at destination. In case of our proposed protocol each RT packet takes round about 0.065ms on average. So, it is clearly seen that transmission delay for RT traffic is greatly reduced when queuing delay is reduced, which is required for RT traffic over network. Finally, it is proved from the performance analysis that the proposed improved AODV routing protocol for real time traffic shows better results than basic AODV.One thing is noted that the Best effort traffic is not badly affected which is given smaller transmission opportunity as compare to the RT traffic. BE packets are now not got stocked in queue for longer than usual because of balanced BE traffic aware rout discovery mechanism. Although, our scope is RT traffic as proposed method is defined to support QoS for real time traffic over MANET.

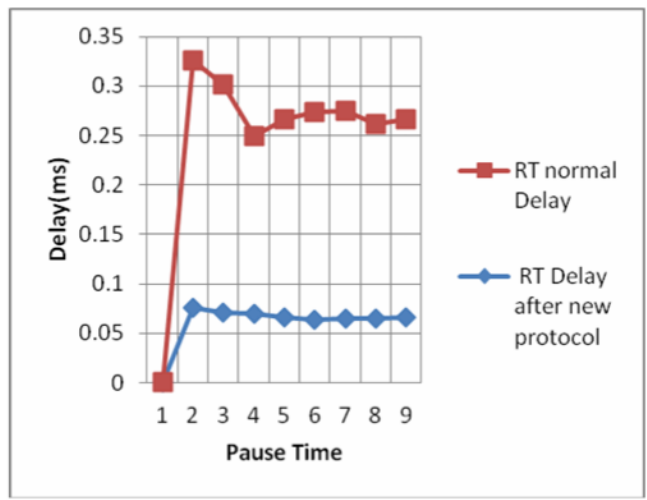

Figure 7. Delays vs. Pause Time

## 5. CONCLUSIONS

We proposed a protocol for real time traffic over the Mobile ad hoc networks by modifying basic AODV. The rout discovery mechanism of basic AODV is modified and BE traffic aware rout discovery mechanism is proposed. New rout discovery mechanism selects routs for transmission that have lesser BE traffic on it. Another procedure is incorporated in basic AODV that give more transmission opportunity for Real Time traffic in MANET. The proposed procedure assigns more bandwidth to RT traffic and reduces transmission delay for RT data packets without effecting Best Effort traffic. Now a RT data packet takes much less time to reach at destination as it requires for efficient transmission. After the implementation the proposed protocol shows greater improvement in throughput and reduces delay in greater extent as well. Hence, for the transmission of real time data over the MANET of moderate density and mobility the proposed improved QoS protocol is more suitable.